# Free-Running Time-Resolved First-Pass Myocardial Perfusion Using a Multi-Scale Dynamics Decomposition: CMR-MOTUS

Thomas E. Olausson[1,2], Maarten L. Terpstra[1,2], Niek R.F. Huttinga[1,2], Casper Beijst[2], Niels Blanken[3], Teresa Correia[4,5], Dominika Suchá[3], Birgitta K. Velthuis[3], Cornelis A.T. van den Berg[1,2], and Alessandro Sbrizzi[1,2]

[1]Computational Imaging Group for MR Therapy and Diagnostics, University Medical Center Utrecht, Utrecht, Netherlands,

[2]UMC Utrecht Cancer Center, Department of Radiotherapy, University Medical Center Utrecht, Utrecht, Netherlands,

[3]Department of Radiology, University Medical Center Utrecht, Utrecht, Netherlands,

[4]School of Biomedical Engineering and Imaging Sciences, King's College London, London, United Kingdom,

[5]Centre for Marine Sciences (CCMAR), Faro, Portugal

Version 3, 2024-10-28

## Abstract

We present a novel approach for the reconstruction of time-resolved free-running first-pass myocardial perfusion MRI, named CMR-MOTUS. This method builds upon the MR-MOTUS framework and addresses the challenges of a contrast varying reference image. By integrating a low-rank plus sparse (L+S) decomposition, CMR-MOTUS efficiently captures both motion fields and contrast variations. This innovative technique eliminates the need for acquiring a motion-static reference image prior to the examination, thereby reducing examination time and complexity for cardiac MRI examinations. Our results demonstrate that CMR-MOTUS can successfully disentangle different dynamic components, offering high-quality motion fields and motion correct a myocardial first-pass perfusion image series.

## Introduction

First-pass myocardial perfusion with magnetic resonance imaging (MRI) is primarily used to detect ischemia in patients with (suspected) coronary artery disease (CAD)[1,2]. These examinations involve a continuous acquisition of at least 60 seconds spanning several types of dynamics, including cardiac motion, respiratory motion, bulk motion, and gadolinium-based contrast agent (GBCA) inflow. To accurately quantify the initial inflow of the GBCA, high spatiotemporal resolution imaging must be obtained during this narrow temporal window, which makes the examination time-sensitive. Hence, the motion, anatomy, and contrast dynamics need to be separated when reconstructing time-resolved MRI to have an accurate first-pass myocardial perfusion quantification. Additionally, the extracted motion dynamics can be used for explicit motion quantification which has clear benefits: high-quality motion fields are diagnostically valuable as they can be used in further image analysis, such as characterizing myocardial strain[3]. Therefore, efficient time-resolved reconstruction methods that enable the separation of motion and contrast dynamics are needed for accurate first-pass myocardial perfusion.

In cardiovascular MRI (CMR), numerous studies have focused on enhancing the efficiency and accuracy of image reconstruction by addressing both cardiac and respiratory motions within a free-running CMR protocol[4,5]. These methods also work well when incorporating explicit signal evolution knowledge, such as for quantitative T1/T2 mapping MRI. However, incorporating the signal evolution during first-pass myocardial perfusion from the GBCA is challenging. Recently developed methods attempted to resolve bulk and respiratory motion using Robust principal component analysis (RPCA)[6] techniques on electrocardiogram (ECG)-triggered first-pass myocardial perfusion to disentangle the contrast enhancement from the registration.[7,8] Other methods have looked into ECG-free first-pass myocardial perfusion examinations by extracting a surrogate signal for the cardiac phase during a continuous acquisition followed by sorting the data based on the surrogate signal before motion correcting for the respiratory and bulk motion in a cardiac phase[9]. However, these registration algorithms struggle with the strong intensity variations during the GBCA inflow, which make cardiac motion correction extremely difficult[10]. Finally, methods that rely on sorting the data into cardiac and respiratory phases or triggering are inherently time-inefficient and sensitive to bulk motion, which can corrupt the time-sensitive inflow of the contrast agent[11].

To use all the data efficiently and reconstruct time-resolved MRI, researchers have focused on integrating powerful methods from video processing, such as the low-rank plus sparse decomposition (L+S)[12,13]. The low-rank (L) component generally captures changes in larger structures that are present, while the sparse (S) component usually captures the small-scale intensity variations such as the enhancement of vessels[12]. Overall, L+S is a flexible regularization method representing spatiotemporal dynamics using simple basis functions, constraining the solution space. However, capturing complex dynamics requires many principal components to obtain an accurate representation, limiting the effectiveness of this method. Alternatively, motion dynamics are more compressible with respect to degrees of freedom than image dynamics and better suited for these applications[14]. Additionally, the inclusion of motion fields into the image reconstruction problem has been shown to reduce blurring in motion-guided L+S reconstructions[15,16]. Hence, this motivates the need to extract high-quality time-resolved motion fields.

One method explicitly relating the k-space data of frames to low dimensional motion fields of a high spatial resolution motion-static reference image, dubbed MR-MOTUS, has demonstrated promising results in extracting high-quality breathing motion fields[17,18]. This approach facilitates time-resolved reconstructions with high temporal resolutions of 100 ms in free-running 3D cine MRI by separating respiratory and bulk motion[17]. MR-MOTUS' success relies on the assumption that a dynamic object in an MRI scanner can be described as a single, static reference image that is deformed by low dimensional motion fields, implying the conservation of magnetization during deformation. This assumption is valid in abdominal imaging but is not valid in cardiac imaging as blood and GBCA flow in and out of the field of view. Another technical drawback of MR-MOTUS is that it requires a motion state reference image before the motion estimation step. Obtaining a reference image may not always be feasible or easily accessible. While breath-holds or gating techniques can be used to obtain these reference images[18], they can prolong the examination time and pose extra challenges for clinicians.

In this work, we investigate the reconstruction of time-resolved free-running first-pass myocardial perfusion MRI using a **C**ontrast-varying **M**odel-based **R**econstruction of **Mot**ion from **U**ndersampled **S**ignal (CMR-MOTUS) that extends the original MR-MOTUS [17–19] framework by reconstructing, in addition to the motion fields, an L+ S[12,16] image series. Cardiac, respiratory, and bulk motion dynamics are extracted using MR-MOTUS whilst the GBCA dynamics are captured using an L+S decomposition. Thus, we efficiently disentangle the different types of dynamics during a time-resolved free-running first-pass myocardial perfusion acquisition from patient datasets. This approach offers significant advantages by reducing examination time and complexity while providing high-quality motion fields and motion corrected perfusion image series.

# Theory

In a conventional MR-MOTUS framework, we define target time-series images $H = [H_1, \dots, H_M]$, a motion static reference image $q$ and motion fields $D = [D_1, \dots, D_M]$ for each dynamic index $t$, up to $M$ total number of dynamics. These quantities are mutually related as follows,

$$H_t(r) = q(D_t(r)) \cdot \det(\nabla D_t), \qquad (1)$$

where $r = (x, y, z)$ are spatial coordinates and $\nabla$ is the Jacobian operator. From equation (1), $H_t$ is approximated by warping the reference image $q$, which is the fundamental requirement for the MR-MOTUS framework. This approximation is incorporated into the MR signal model leading to the following MR-MOTUS signal model $\widehat{F}$ [18].

$$s_t(k) = \widehat{F}(D_t|q) = \int_\Omega q(r) \, e^{-i2\pi k \cdot D_t(r)} dr. \qquad (2)$$

Here, the reference image $q$ is related to the motion fields $D$ at time point $t$ for the measured k-space data $s_t$, and $k = (k_x, k_y, k_z)$ are k-space coordinates.

One limitation of this approach is that it does not consider contrast variation in the reference image which can occur in cardiac examinations due to blood flow or contrast administration. Additionally, $q$ must be motion static, meaning that breath-hold or gated images need to be acquired in a preparation phase before the examination to use the MR-MOTUS framework. To tackle these challenges, we made a small but crucial extension to the MR-MOTUS framework where the reference image is allowed to change over time, resulting in a time series of images, $Q = [q_1, \dots, q_M]$. Although this modification enhances the validity and flexibility of the MR-MOTUS model, it also drastically augments the number of unknowns in the reconstruction problem, making it even more ill-posed and less challenging to solve for the motion fields. To correct for this, we use the low rank plus sparse (L+S) decomposition[20] as a regularizer for $Q$. We call this joint approach CMR-MOTUS. Additionally, there is no longer a need for a preparation phase to acquire a motion static reference image, instead contrast-varying motion static reference images are continuously estimated for during alternating iterations using the entire time-resolved data.

Thereby, $H$ is then approximated by,

$$H_t(r) = q_t(D_t(r)) \cdot \det(\nabla D_t), \qquad (3)$$

Hence a slightly modified signal model $F$ is derived to include this time-varying reference image $Q$

$$s_t(k) = F(D_t|q_t) = \int_\Omega q_t(r) \, e^{-i2\pi k \cdot D_t(r)} dr. \qquad (4)$$

Even in the extended model, there is still a high chance that motion will be captured by the L+S decomposition of Q and therefore corrupt its anatomical information. Moreover, contrast variation will be captured by the motion fields in the MR-MOTUS framework which corrupts their quality. The goal is to gradually disentangle the motion and contrast variations by alternating between the L+S decompositions and the MR-MOTUS framework.

In summary, CMR-MOTUS relies on the alternating solution of two reconstruction problems, namely:

1. **Image Reconstruction** to determine the time-dependent image contrast changes $Q$.
2. **Motion Estimation** to determine the motion fields $D$.

We briefly describe both alternating steps.

## 1. Image Reconstruction

In the L+S decomposition of time-resolved MRI, the entire target time-series images $H$ is approximated by a low rank term $L$ plus a sparse term $S$. The $L$ term captures the larger scale slow changes in the time-series, while the $S$ term captures the finer details in the series, including the contrast enhancement[20]. For implicit low-rank modeling in time-resolved MRI reconstruction, the following L+S inverse problem[20] is solved

$$[L, S] = \arg\min_{L,S} \sum_{t}^{M} \|B(H_t) - s_t\|_2^2 + \lambda_L \|L\|_* + \lambda_S \|S\|_1 \quad \text{with } H = L + S, \tag{5}$$

where $B$ denotes the combined action of common linear MRI encoding operators such as the Fourier transform, sampling mask and coil sensitivity weighting. Furthermore, $\|\ \|_*$ denotes the nuclear norm, $\|\ \|_1$ the L1-norm, and $\lambda_L$ and $\lambda_S$ are the weights for the regularization terms involving $L$ and $S$, respectively.

We leverage motion fields using MR-MOTUS to motion correct the L+S reconstruction process. This is achieved by substituting the data consistency term in eq (5) with the CMR-MOTUS model presented in eq (4). The final L+S inverse problem becomes

$$[L*, S*] = \arg\min_{L,S} \sum_{t}^{M} \|F(q_t|D_t) - s_t\|_2^2 + \lambda_L \|L\|_* + \lambda_S \|\mathcal{T}S\|_1 \quad \text{with } Q = L + S, \tag{6}$$

where $Q = L + S$ now approximates the time-varying reference image, and the encoding operator now contains time-dependent motion fields as described by $F$, sampling mask and coil sensitivity weighting. Additionally, we take advantage of sparsity in the temporal Fourier domain using the temporal Fourier operator $\mathcal{T}$, as it has shown promising results in first-pass myocardial perfusion applications[20].

Technically, we have the flexibility to designate any motion state at time $t$ as a reference motion static state by subtracting its influence on all the other motion fields. For instance, the reference static phase could correspond to the diastolic one. This adjustment is done before equation 6 and it steers the L+S reconstruction process towards a diastolic reference motion static state. For all datasets we process we opt to reconstruct a mid-position $Q$ by subtracting the temporal mean of the motion fields from $D$.

Since we use a non-differentiable regularizer, we use a custom proximal gradient method based on the Fast Iterative/Shrinkage Thresholding Algorithm (FISTA)[21,22] to solve equation (6).

## 2. Motion Estimation

As described in Huttinga et al[17], non-rigid motion fields in MRI can be decomposed into a low rank model with principal components $\Phi$ and $\Psi$. For an explicit Rank $R$, $\Phi$ is a matrix with size $N \times R$ describing the spatial components and $\Psi$ is a matrix with size $N \times M$ describing the temporal components. The full motion fields are then approximated by $D \approx \Phi\Psi^\top$. Additionally, we parameterize the low-rank motion fields using a B-spline basis in space and time as described by Huttinga et al[17]. Hence, we can solve the following inverse problem for an explicit low-rank representation of the motion fields, similar to the Low Rank MR-MOTUS method, but employing a time-varying contrast image $q_t$ rather than a fixed reference:

$$[\mathbf{\Phi}^*, \mathbf{\Psi}^*] = \arg\min_{\mathbf{\Phi}\mathbf{\Psi}^\top} \sum_t^M \|F(\mathbf{D}_t|\mathbf{q}_t) - \mathbf{s}_t\|_2^2 + \lambda_{\text{TV}} \|\text{TV}(\mathbf{D})\|_{2,1}. \quad (7)$$

This inversion problem is regularized with a vectorial total variation operator TV which penalizes the $L^2$-norm over the total variation per motion-field direction[23]. We use the L-BFGS[24] algorithm to solve equation (7).

Figure 1 summarizes the CMR-MOTUS framework. The iterative process alternates between steps 1 and 2 until the time-varying reference image of the motion state is accurately captured in step 1 and the motion dynamics are effectively captured by the motion fields in step 2. The optimal regularization weights $\lambda_L$, $\lambda_S$, $\lambda_{\text{TV}}$, and the maximum number of alternations and inner iterations are determined empirically (see Methods section).

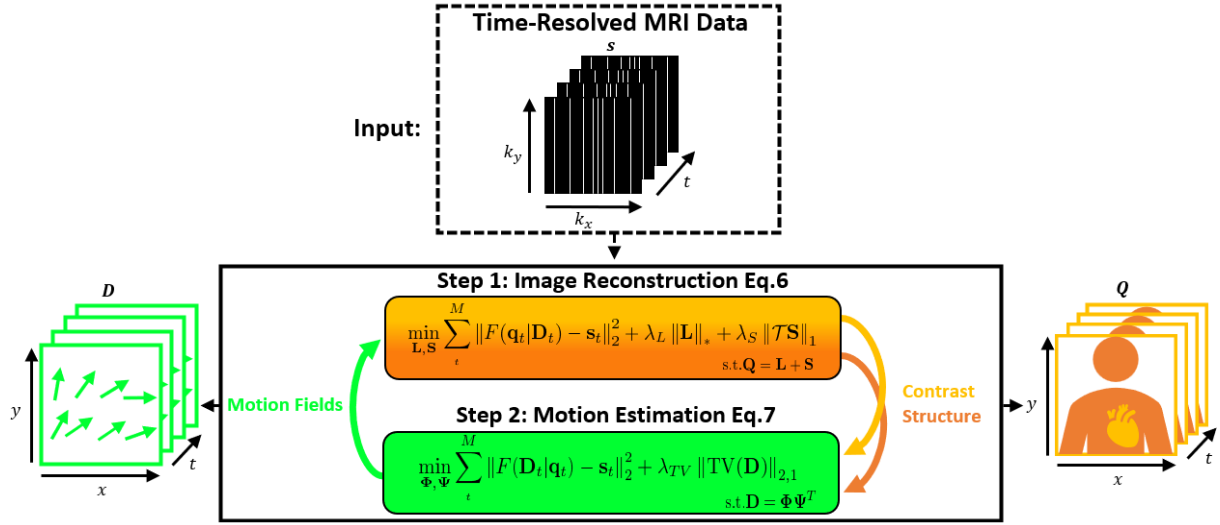

*Figure 1: Schematic representation of the proposed CMR-MOTUS framework. The framework takes as input (top) raw time-resolved t MRI data $s_t$. Step 1 is where the contrast-varying reference images $q_t$ are solved for while the motion fields D are fixed. This is done using a L+S reconstruction. Then, in step 2, the motion fields D are solved for while q is fixed. The output of step 1 is shown on the right as the contrast component S and structure component L. Step 1 and step 2 are solved iteratively in an alternating manner until the motion and contrast are disentangled. The output of step 2 is shown on the left as the motion fields $D$.*

## Methods

First, we test our proposed CMR-MOTUS method on first-pass myocardial perfusion simulations using the MRXCAT[25] framework. This is to show that we achieve similar image quality of our motion-resolved cines compared to compressed sensing whilst extracting diagnostically valuable motion fields.

Subsequently, we demonstrate the feasibility of the proposed method on two publicly available datasets of cardiac cine MRI without contrast enhancement taken from the OCMR dataset[26]. They provide in-vivo prospectively undersampled raw cartesian multi-coil k-space CINE data, which were acquired during free-breathing and without ECG-triggering. This data is affected by both breathing and cardiac motion dynamics, which we resolve in the proposed reconstruction. Ground truth reconstructions are not available. Hence, we validate our reconstructions with quantitative results using the Tenengrad function, a gradient-based sharpness metric of the motion-resolved image series $H$.

Finally, we use two first-pass myocardial perfusion MRI datasets of patients prospectively scheduled for a cardiac MRI examination to analyze the proposed CMR-MOTUS reconstruction method. The first dataset was acquired with ECG-triggering and a breath hold whilst the other dataset was acquired without ECG-triggering and in free breathing. These first-pass myocardial perfusion reconstructions will highlight the ability of the proposed framework to separate the physiological motion dynamics from the contrast agent inflow.

Implementation

The proposed alternating reconstruction is a custom extension to the Low Rank MR-MOTUS[17] framework together with an inhouse L+S decomposition implementation in MATLAB. All reconstructions are performed on the CPU of a computational server (AMD Ryzen Threadripper 3970X 32-Core 3700 MHz and 256 GB RAM). The number of alternating steps is decided empirically (convergence was usually observed after 5 alternating steps), with a maximum number of inner iterations set to be 60 in step 2, and step 1 is set to break if the normalized difference in the total objective value was below $1 \cdot 10^{-4}$ over 4 consecutive iterations. Step 1 usually converged after 10 iterations at early alternating steps.

The entire proposed alternating framework, which illustrated in Figure 1, is initialized with a time-averaged reference image generated by step 1 with $\mathbf{D} = 0$. Then, step 2 begins with this reference image. At each alternating of step 1, $\mathbf{\Phi}$ and $\mathbf{\Psi}$ are initialized with random numbers between $-1$ and $1$. We found that the first estimated motion fields. Whilst in step 2, the L+S reconstructions were initialized with the previous estimations of the $\mathbf{L}$ and $\mathbf{S}$ components. The step size for step 2 is determined to be $1.3/\gamma$ [22], where γ is the largest eigenvalue of $\mathbf{FF}^*$ which is determined via a power method[27].

Test 1: In-silica

The XCAT phantom was used to generate six hundred frames with cardiac motion at a temporal-resolution of 50 ms and a matrix size $100 \times 100$. The orientation is set to a short-axis view of the heart. The heart cycle period was set to 1 s. With these settings, XCAT provided ground truth (GT) segmentations and motion fields for each time frame.

A saturated-recovery gradient echo sequence of first-pass myocardial perfusion was simulated resulting in coil sensitive MR-contrast XCAT frames using the MRXCAT[25]. The acquisition parameters were $T_R = 2.0$ ms, $T_{sat} = 150$ ms, flip angle $= 15°$, and 8 number of coils. We simulated undersampling by multiplying a sampling mask with the Fourier transform of each frame. The sampling mask followed a pseudo-random Cartesian sampling pattern[28] with 12 readouts per frame, corresponding to an acceleration factor of 10. Coil maps were estimated using ESPIRT[29]

Reconstructions
- **Proposed (CMR-MOTUS):** We performed the proposed alternating reconstructions on the in-silica data with $\lambda_S = 10$ and explicitly enforce rank one for both the motion fields and the images $\mathbf{L}$ component with the aim to reconstruct only cardiac motion fields and a static single frame low-rank image. The motion model used 80 b-spline control points in each spatial direction. Regularization weights were set to $\lambda_{TV} = 1 \cdot 10^{-8}$ and $\lambda_J = 10$.
- **Compressed Sensing:** For comparison purposes, a compressed sensing (CS) reconstruction was performed on the same data with the BART[30] toolbox using empirically determined spatial wavelet and temporal total variation regularization weights set to $\lambda_{wavelet} = 0.001$ and $\lambda_{total\ variation} = 0.6$ respectively. The motion fields of the CS reconstruction were separately determined for each reconstructed frame with the Elastix[29] registration toolbox using the first frame as the fixed image.

## Quantitative Validations

- **Image Quality:** The mean structural similarity index (SSIM)[32] with respect to the ground truth MRXCAT frames was determined for the CMR-MOTUS time-series images **I**, and the CS reconstruction.
- **Motion fields:**
  The end point error (EPE) with respect to ground truth XCAT motion fields over the entire field of view (FOV) was determined for the proposed MR-MOTUS motion fields, and the motion fields estimated with Elastix after CS reconstruction. To evaluate the accuracy of the motion fields over the myocardium, the first segmentation of the XCAT series was warped according to both methods, which resulted in two time-resolved segmentations. The DICE score with the XCAT segmentations as a reference was determined for the warped first frame XCAT segmentation of the myocardium with the estimated motion fields from MR-MOTUS and Elastix respectively.

## Test 2: In-vivo free-running time-resolved CMR

Two publicly available multi-coil k-space datasets of prospectively undersampled 2D free-breathing balanced steady state free precession (bSFFP) single-slice cardiac MRI cine series in a short axis (SAX) and long axis (LAX) view were used in the analysis[26]. Both datasets were reconstructed using the proposed CMR-MOTUS reconstruction and a CS reconstruction. The datasets acquired 3 and 6 consecutive heart cycles during free-breathing (time-resolved) which resulted in 65 and 128 temporal frames, respectively. Coil maps were estimated with ESPIRIT. We refer the reader to the reference for more details on the acquisition.

For the proposed CMR-MOTUS reconstruction of each dataset, we explicitly model one component in the **L** images and used $\lambda_S = 1$. The motion model was explicitly defined to use 4 components (Rank=4) and used 90 b-spline control points in each spatial direction and 40 in the temporal domain. Regularization parameters were set to $\lambda_{TV} = 1 \cdot 10^{-8}$. For comparison, a CS reconstruction was performed using the BART toolbox using spatial wavelet and temporal total variation regularizations set to $\lambda_{wavelet} = 1 \cdot 10^{-6}$ and $\lambda_{total\ variation} = 1 \cdot 10^{-6}$, respectively. Since Ground Truth data is not available, the Tenengrad sharpness metric was determined to evaluate image quality for the proposed CMR-MOTUS time-series images **I** and the CS reconstruction.

## Test 3: In-vivo breath-hold ECG-triggered First-Pass Myocardial Perfusion

One patient dataset of an ECG-triggered saturation-recovery TFE sequence during first-pass myocardial perfusion was used in the analysis to showcase bulk motion correction. The data was acquired on a Philips 3T Achieva scanner. In-plane resolution was $1.6 \times 1.6\ \text{mm}^2$, FOV = $320 \times 320\ \text{mm}^2$, slice thickness = 10 mm, 32 coils, and a total acquisition time of 1 minute. A radial sampling scheme was used with 10 spokes per frame. The k-space was gridded to a Cartesian grid. Coil maps were estimated using ESPIRIT. The patient's breath-hold was imperfect resulting in bulk motion as the patient inhaled again. This dataset was used to showcase how the proposed CMR-MOTUS framework will perform with imperfect breath-holding during first-pass myocardial perfusion.

A standard L+S (equation 5) and CMR-MOTUS reconstruction were performed to visually compare the result of no motion correction and motion correction. For both reconstructions, $\lambda_S = 1$ and we explicitly model one component in the **L** images. The motion model was explicitly defined to use one component (Rank=1) and used 80 B-spline control points in each spatial direction. No B-spline parameterization of the temporal component was used since temporally non-smooth motion is expected. Motion field regularization was set to $\lambda_{TV} = 1 \cdot 10^{-7}$.

### Test 4: In-vivo free-running time-resolved First-Pass Myocardial Perfusion

Another patient dataset of a free-breathing saturation-recovery bSFFP sequence during first-pass myocardial perfusion without any ECG-triggering was used in the analysis. No patient information was required as only anonymous raw k-space data was used. The data was acquired on a Philips 1.5 T Ingenia scanner. Relevant acquisition parameters include an in-plane resolution of $2.73 \times 2.73$ mm$^2$, FOV = $350 \times 350$ mm$^2$, slice thickness = 10 mm, 16 coils, TE/TR = 1.29/2.6 ms, flip angle = 50°, saturation delay = 100 ms, and a total scan time of 90 s. A pseudo golden angle radial sampling scheme was used with 20 spokes per frame. The k-space was gridded to a Cartesian grid. Coil maps were estimated using ESPIRIT. This data was used to showcase how the proposed CMR-MOTUS framework will perform with contrast enhancement, respiratory motion and cardiac motion.

A standard L+S and the proposed CMR-MOTUS reconstruction were performed to visually compare the result of no motion correction and motion correction. In both frameworks, the **S** component of the images regularization was set to $\lambda_S = 5$ and the **L** component of the images was explicitly set to solve for one component (Rank=1). The motion model was explicitly defined to use R = 2 and used 64 B-spline control points in each spatial direction and 150 in the temporal domain for 800 temporal frames. Regularization parameters were set to $\lambda_{TV} = 1 \cdot 10^{-8}$.

## Results

### Test 1: In-silica

A still frame from the reconstructed cines during an end diastole and contrast inflow is presented in Figure 2. There is less temporal blurring in the CMR-MOTUS cine than in the standard L+S cine as indicated with the blue arrows. This is most apparent where the LV does not contract in the standard L+S cine and the difference is relatively large. Using the MRXCAT cine frames as a reference, the CMR-MOTUS cine and L+S cine resulted in a temporal mean SSIM value of 0.9780 ($SD = 0.0145$) and 0.9706 ($SD = 0.0155$) over the myocardium respectively, Wilcoxon signed ranked test $p < 0.001$. Thus, this indicates an improved image similarity with the inclusion of motion fields in CMR-MOTUS cine. The warped initial myocardium segmentation with the estimated motion fields resulted in a temporal mean DICE score of 0.9117 ($SD = 0.0624$) using CMR-MOTUS and 0.8913 ($SD = 0.0553$) using L+S & Elastix method, respectively, Wilcoxon signed rank test $p < 0.001$. Hence, in this test CMR-MOTUS was able to reconstruct more accurate images as well motion fields w.r.t. state-of-the-art.

Figure 3 showcases the time-resolved EPE of the motion fields estimated with each method with respect to the GT XCAT motion fields over a segmentation of the myocardium. CMR-MOTUS motion fields result in a lower EPE compared to motion fields estimated by Elastix.

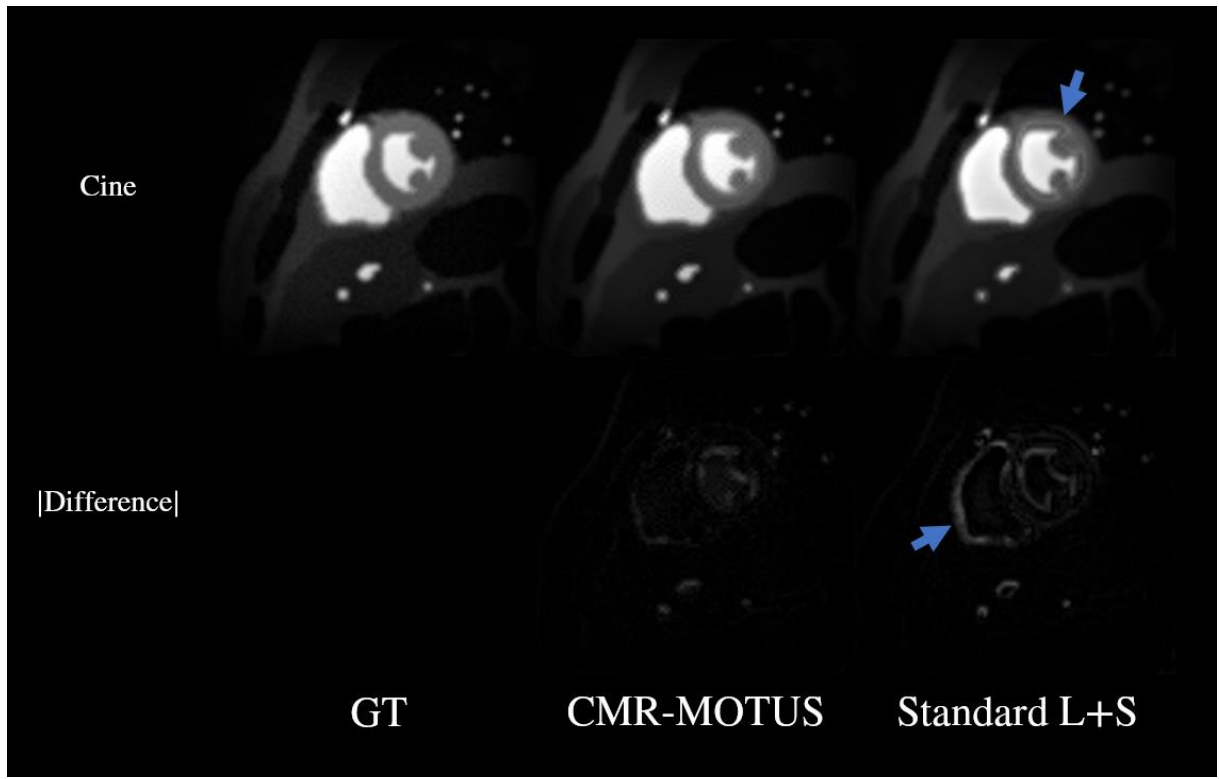

*Figure 2: An end systole frame during contrast agent inflow from the cine reconstructions using data from experiment 1. The left shows the cine reconstructed using a standard low-rank plus sparse (L+S) decomposition. The center shows the cine reconstructed using CMR-MOTUS. The right shows the ground truth (GT) MRXCAT first-pass myocardial perfusion frames.*

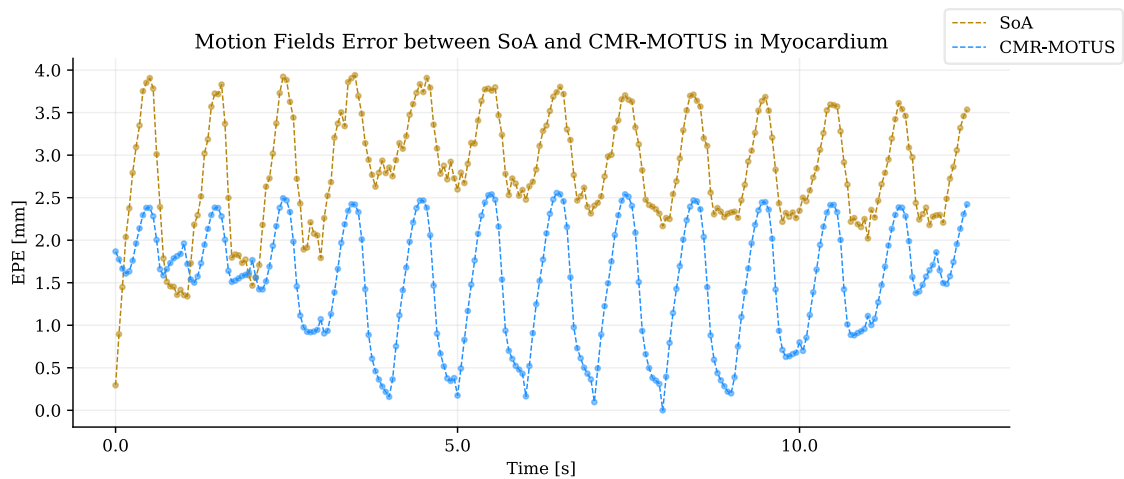

*Figure 3: Data from experiment 1. The plot shows the mean end point error (EPE) in mm as a function of time of the estimated motion fields with the XCAT motion fields as the reference over the entire field of view (FOV). Motion fields were estimated using a combination of low-rank plus sparse (L+S) decomposition image reconstruction then Elastix image registration (gold line), and CMR-MOTUS framework (blue line). There is a clear modulation of the EPE in the CS & Elastix method during the contrast agent inflow whilst CMR-MOTUS is less affected. This indicates a separation of the time-dependent contrast enhancement from the cardiac motion estimation with CMR-MOTUS.*

### Test 2: In-vivo Time-Resolved CMR

The reconstructed SAX cine images are shown in Figure 3. Both methods capture the entire cardiac and respiratory motion dynamics during the acquisition. The mean Tenengrad metric of the CMR-

MOTUS cine was 14% larger than the CS cine, Wilcoxon signed ranked test $p < 0.001$, which indicates a sharper cine is reconstructed when using CMR-MOTUS.

The LAX views in Figure 5 show that both methods capture cardiac and respiratory dynamics. However, this specific LAX view and region contains complex motion dynamics and fast-moving thin structures. Temporal blurring becomes more apparent in these regions in the CS cine than in the CMR-MOTUS cine.

Figure 5 highlights the disentanglement of SAX cine dynamics using CMR-MOTUS. These correspond to the low-rank component of the images ($\mathbf{L}$), the sparse component of the images ($\mathbf{S}$), and the low-rank motion fields ($\mathbf{D}$). Combining these components produces the CMR-MOTUS cine ($\mathbf{I}$). Major anatomical structures are captured in $\mathbf{L}$ in one static motion state, whilst $\mathbf{S}$ captures local intensity changes including blood flow through arteries. There is little residual motion captured in $\mathbf{S}$ which can be due to through-plane motion. From the low-rank motion model, it was possible to separate the cardiac motion fields from the respiratory motion fields based on the frequency content of the temporal scaling components (i.e. the vectors in $\mathbf{\Psi}$ )

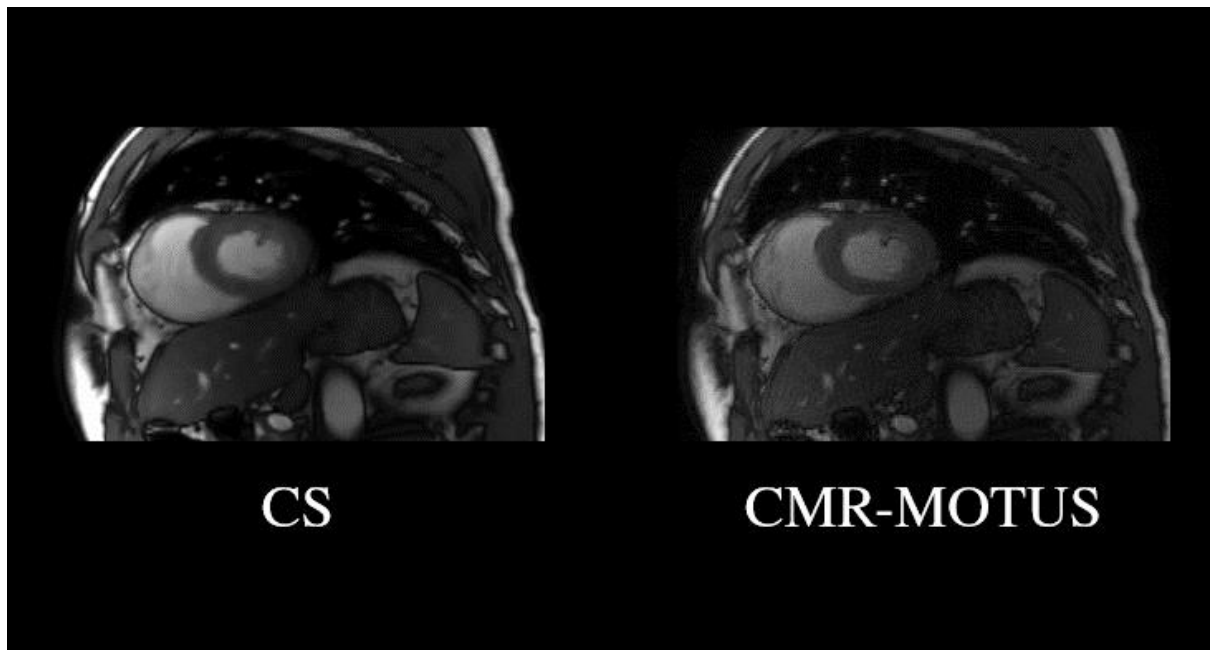

*Figure 4: [animation link for supplementary material] Frame from the reconstructed cines of the short axis (SAX) views in experiment 2 using compressed sensing (CS) on the left and CMR-MOTUS on the right. More anatomical details are resolved in the proposed method than with CS.*

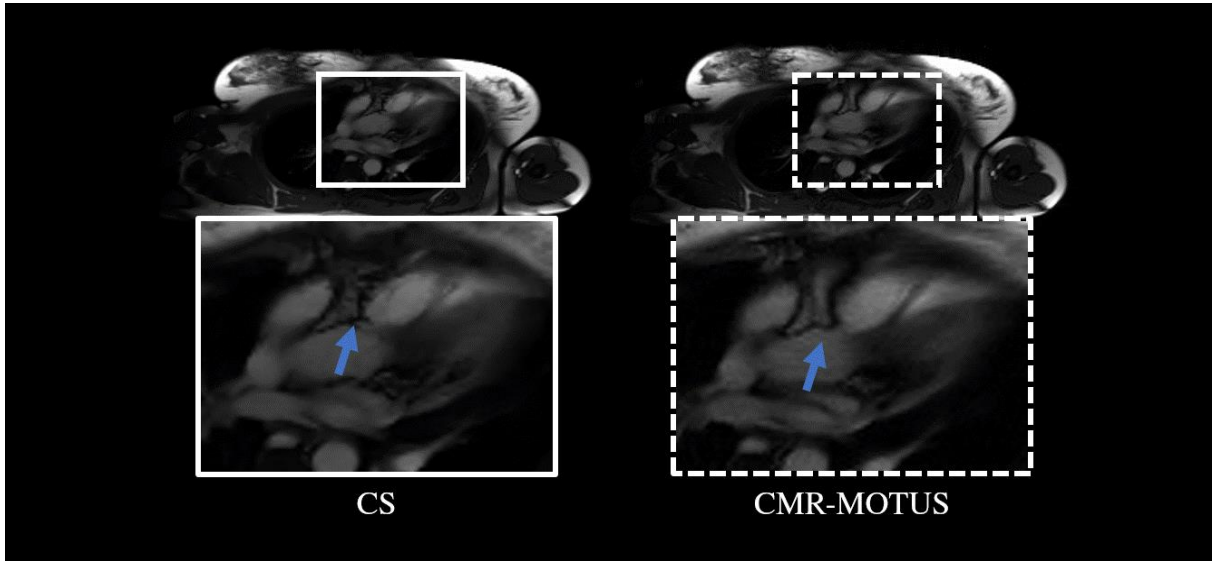

*Figure 5: [animation link for supplementary material] Frame from the reconstructed long-axis (LAX) cine images in experiment 2 using compressed sensing (CS) on the left and CMR-MOTUS on the right. Both methods capture the dynamics in the data. Here the temporal blurring of the CS cine can be seen in the region pointed out with the arrow.*

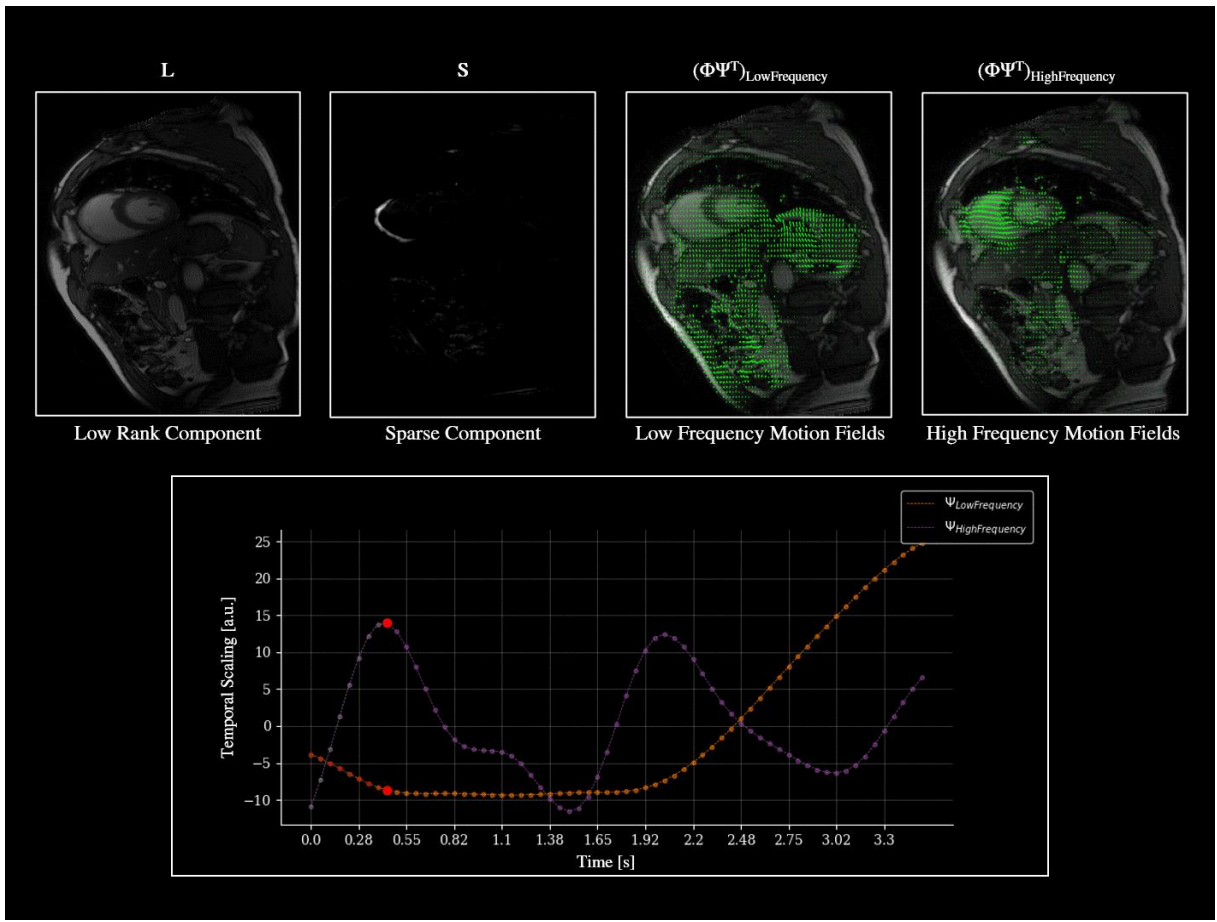

*Figure 6: [animation link for supplementary material] Disentanglement of dynamics in the short axis (SAX) view data in experiment 2 using CMR-MOTUS. Starting above, left to right. The low-rank component captures the major structures in one motion state. The sparse component captures the local intensity changes which are mainly due to residual motion, blood flow, and through-plane motion. The low frequency motion fields are the low-rank motion field components that have a low temporal frequency and these are visualized by warping the **L+S** with the motion fields overlaying. The high frequency motion Fields are the low-rank Motion Field components that have a high temporal frequency, and these are*

*visualized by warping the **L+S** with overlaying motion fields. The temporal scaling Ψ of each low-rank motion field is plotted against time below.*

### Test 3: In-vivo breath-hold ECG-triggered First-Pass Myocardial Perfusion

The reconstructed cines/dynamic images are shown in Figure 7. In the non-corrected **L+S** images, there are two time points where the bulk motion induces blurring, mainly at the edge of the RV indicated with the arrow. CMR-MOTUS can resolve the underlying bulk motion, which is disentangled from the GBCA inflow. Using the motion information, the **L+S** reconstruction can be spatially aligned throughout the entire acquisition resulting in motion-corrected images. The effect of this motion correction is emphasized in the y-t profile plots (Figure 6, right side) as two overshoots (see arrows) from bulk motion are corrected.

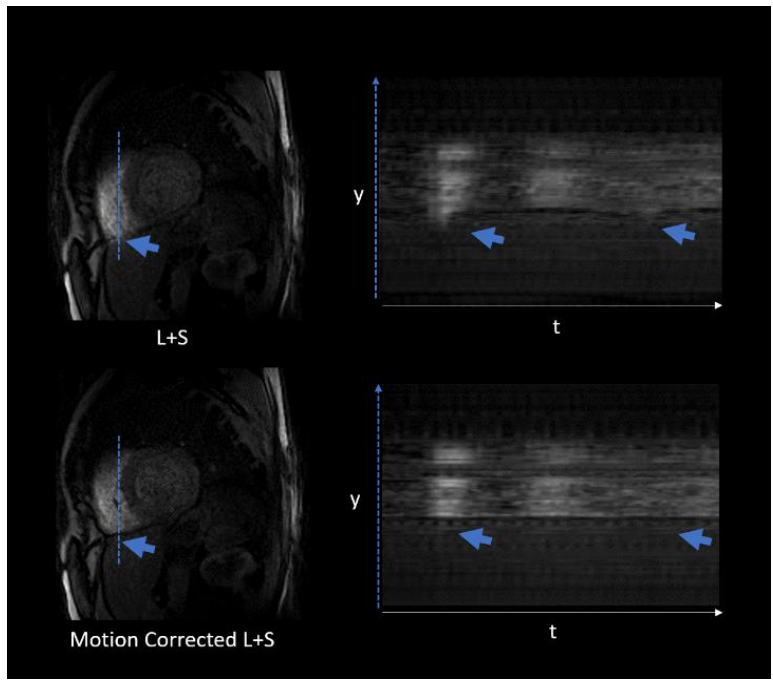

*Figure 7: Reconstructions in experiment 3 of a breath hold (BH) ECG-triggered first-pass myocardial perfusion. The top left shows a frame from the ECG-triggered cine using a standard low rank plus Sparse decomposition (L+S) reconstruction without motion correction. The bottom left shows a frame from the proposed framework (CMR-MOTUS) motion-corrected (L+S) reconstruction using the estimated bulk motion fields. On the right are y-t profiles of the left reconstructions along the dashed lines. Blue arrows point out the locations of motion corruption due to bulk motion from an imperfect breath hold. This bulk motion is corrected as seen in the profile of the Motion Corrected L+S.*

### Test 4: In vivo time-resolved first-pass myocardial perfusion

The reconstructed dynamic images are shown in Figure 8. Both methods capture the motion and contrast dynamics during the entire acquisition. There is temporal blurring at the LV seen in the L+S cine during cardiac contractions, while it is less noticeable in the CMR-MOTUS cine. One possible explanation is that the principal components of the L+S, with these specific parameters (R=1 and strong sparse regularization), are insufficient to describe all the complex contrast variations and motion dynamics present in the data. Motion dynamics are more compressible than image dynamics, hence the addition of a low-rank motion field assumption into the reconstruction reduces the temporal blurring of the images.

Figure 9 highlights the disentanglement of the motion components. Major anatomical structures are captured in **L** in one static motion state, while **S** captures the contrast dynamics during first-pass perfusion. There is little residual motion captured in **S** which can be due to through-plane motion. From the low-rank motion model **D** it was possible to separate different types of motion based on the frequency content of the temporal scaling components $\Psi$ and direction of the principal spatial components $\Phi$. For the high temporal frequency motion fields, the corresponding spatial components have different directions within the myocardium whilst the low temporal frequency motion fields have a single direction (feet-head) within the myocardium. This can indicate that the high-frequency motion fields capture cardiac contractions and the low-frequency motion fields capture the almost bulk-like breathing motion.

Leveraging the disentanglement of motion dynamics, a motion-corrected L+S reconstruction was achieved using all the data from the entire acquisition. Figure 9 shows the x-t and y-t plot with and without this motion correction in the L+S reconstruction. Modulations due to respiratory and cardiac motion are seen in the x-t and y-t plot without motion correction. These modulations are not seen in the motion corrected plots, hence giving a motion corrected GBCA inflow.

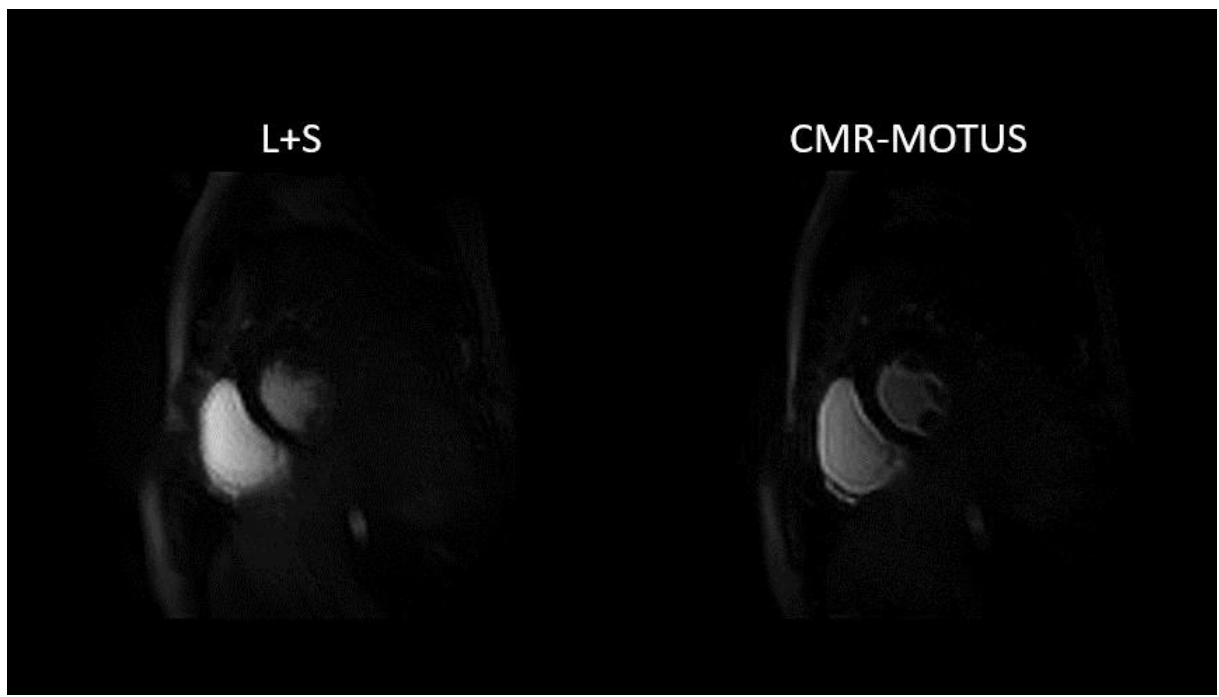

*Figure 8 [animation link for supplementary information]: Cine reconstructions using data in experiment 4 of free-breathing first-pass myocardial perfusion without ECG-triggering. Left shows a standard low rank plus Sparse decomposition (L+S) reconstruction and right shows the CMR-MOTUS cine. There is more blurring seen in the standard L+S reconstruction whilst the CMR-MOTUS cine seems smoother in motion.*

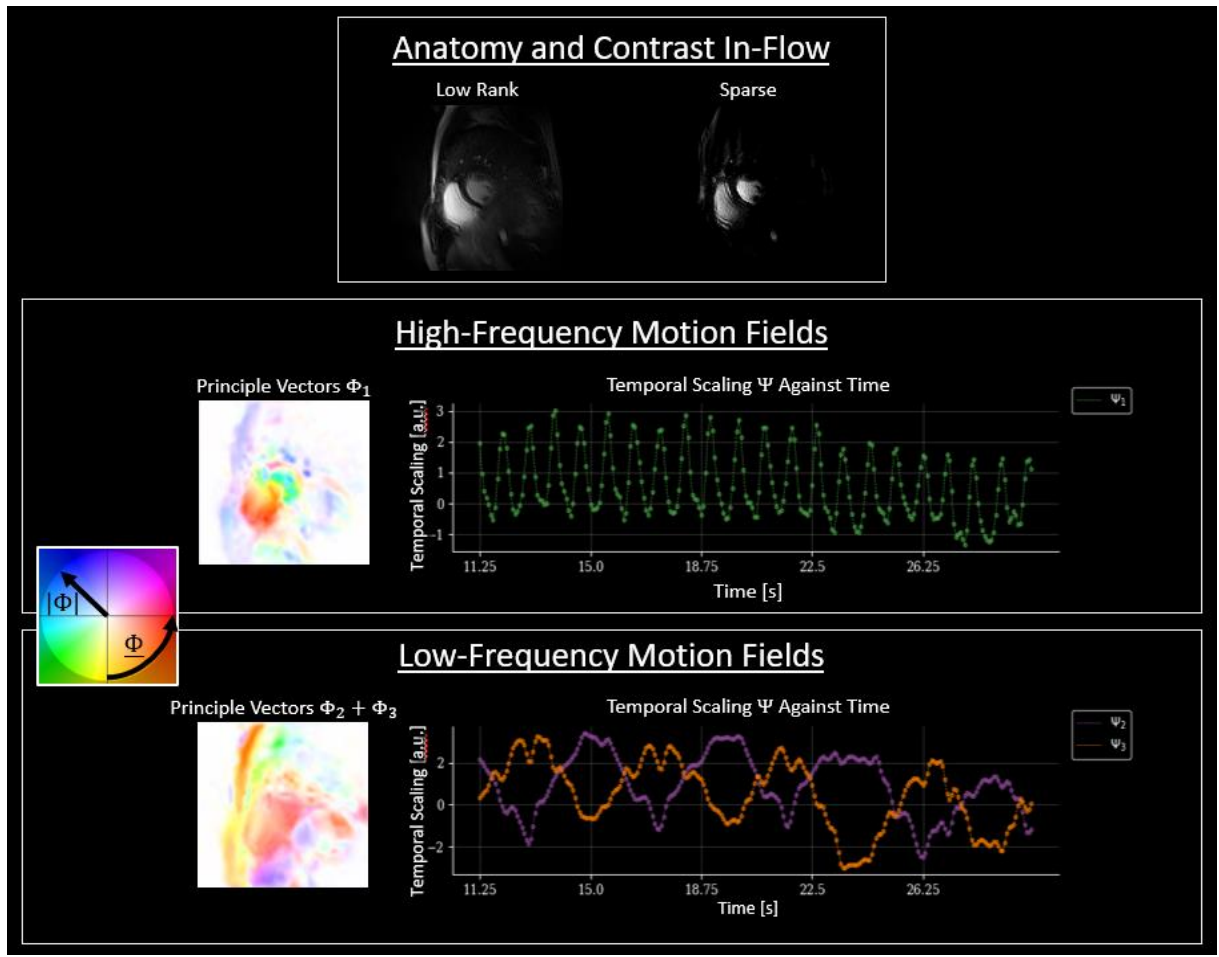

*Figure 9: CMR-MOTUS dynamic disentanglement using data in experiment 4 of free-breathing first-pass myocardial perfusion without ECG-triggering. Top shows the components which capture the anatomy and the contrast in-flow, which are the low rank component and sparse component, respectively. Middle shows the high-frequency motion fields with their principal vectors $\Phi_1$ visualized on the left and their temporal component $\Psi_1$ plotted against time on the right. Similarly, the bottom shows the high-frequency motion fields with their principal vectors $\Phi_2 + \Phi_3$ visualized on the left and their temporal component $\Psi_2 + \Psi_3$ plotted against time on the right.*

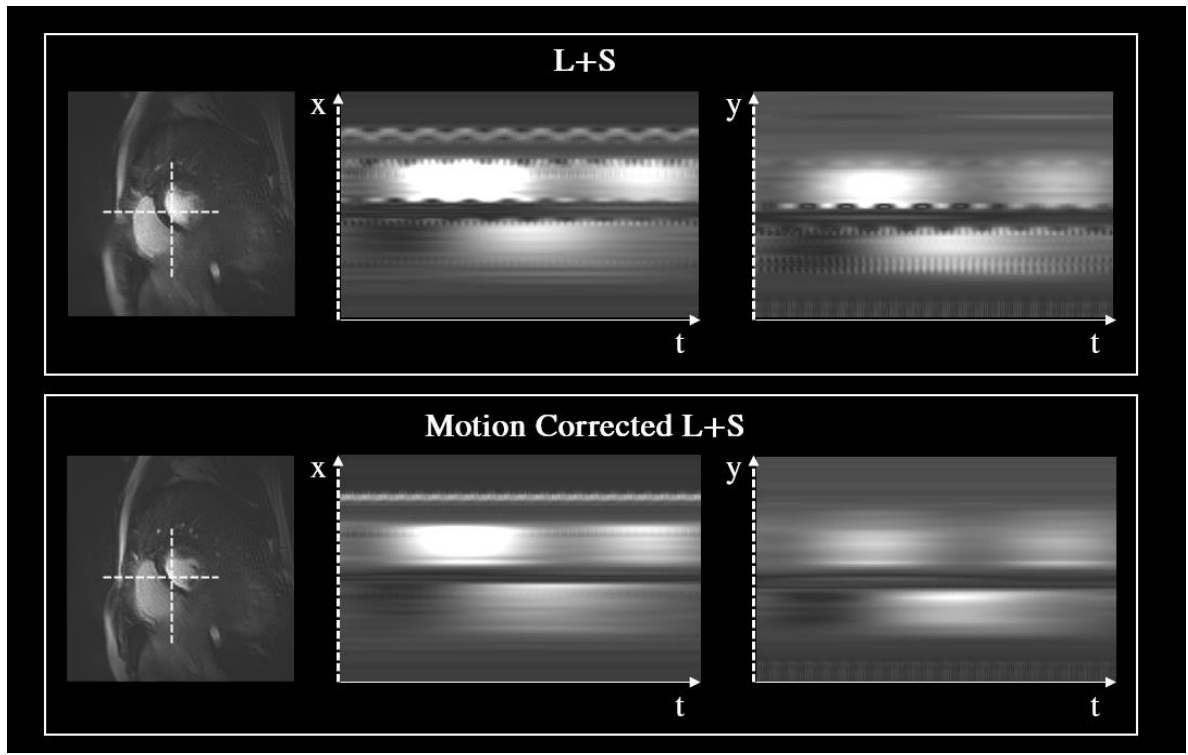

*Figure 10: x-t profile plots obtained using data from experiment 4 of free-breathing first-pass myocardial perfusion without ECG-triggering. (Top) CMR-MOTUS motion corrected low rank plus sparse decomposition (L+S) reconstruction, x-t and y-t plot (Bottom) Standard L+S reconstruction, x-t and y-t plot. (Left) short-axis (SAX) image indicates where the profile was drawn (dashed line). It is clear that the standard L+S x-t plot is modulated by respiratory and cardiac motion. The proposed L+S x-plot shows that the anatomy is spatially aligned since the motion is corrected for.*

## Discussion

We presented a multi-scale dynamic decomposition framework to capture cardiac/respiratory motion and contrast dynamics in first-pass myocardial perfusion MRI without ECG-gating and breath holds. In contrast to traditional methods which require ECG triggering of CMR, the method presented enabled a more efficient acquisition by using a time-resolved acquisition which correlates all timepoints through a motion model. This was achieved using a joint-reconstruction of high-quality motion fields with the extended MR-MOTUS model and a motion-corrected L+S image model. This combination of frameworks enabled the disentanglement of different motion components such as respiratory and cardiac motion.

To successfully achieve our goal, we must overcome a limitation of the previous MR-MOTUS implementation: the reference image was assumed to be static, meaning no changes in contrast or external signal sources. This assumption fell short in describing mass flowing into the FOV, especially problematic in cardiac MRI examinations where there is frequent blood inflow and outflow. The proposed contrast time-varying reference image gives more flexibility to the model to describe this mass inflow. Although we do not yet have empirical results explicitly demonstrating this added flexibility, the theoretical framework and other results suggest a significant improvement in modeling physiological contrast and motion dynamics.

One common method of mitigating motion artifacts is to discard parts of the data which contain bulk motion. However, this might not be feasible in time-sensitive examinations such as in first-pass myocardial perfusion. This is seen in figure 6 (top row, pink arrows) where images of contrast inflow are corrupted due to an imperfect breath hold, which cannot be discarded for perfusion analysis. With CMR-MOTUS, the bulk motion is estimated and is corrected for in the image reconstruction as seen in figure 6 (bottom row, pink arrows). Another motion mitigation method triggers scans to a certain phase in the breathing or cardiac cycle with help from breath holds and ECG, which effectively reduces the efficiency of the scan. However, irregular heart rates impact the temporal resolution negatively as improving the temporal resolution of the scan improves the downstream task of perfusion quantification[33]. With CMR-MOTUS, motion is mitigated through the estimation of high-quality motion fields to use a free-running continuous data-stream as seen in figure 9.

Multi-resolution registration has shown promising results in applications to cardiac MRI[34]. The current implementation of CMR-MOTUS uses the same resolution for motion fields across each motion field component. However, the scale of motion may vary depending on the source, with cardiac motion potentially being smaller than respiratory motion. Using a multi-resolution motion model is worth exploring. In addition to a non-rigid motion model, including an affine motion model to capture bulk motion would also be interesting to explore.

The use of a multi-scale low-rank decomposition in dynamic contrast-enhancement MRI has also shown promising results[11]. This multi-scale approach enabled the reconstruction of high temporal resolution 3D dynamic MRI data which captured breathing motion and contrast inflow. Additionally, it enabled a memory efficient reconstruction of this large-scale 3d+time problem by explicitly solving for a selected number of multi-scale low-rank components and using a stochastic gradient decent algorithm. An approach similar to this with the inclusion of multi-resolution low-rank motion fields could reduce the number of unknowns of the reconstruction problem and is worth exploring in the future.

The previous work on MR-MOTUS has shown that low spatial frequency data is sufficient to reconstruct motion fields[18]. Furthermore, deep learning methods have shown that the central regions of k-space play an important role in the estimation of non-rigid motion fields[35]. Other studies have shown that contrast enhanced images can be reconstructed by acquiring low frequency regions in k-space and borrowing pre-contrast enhanced high frequency k-space acquired with the same anatomy and FOV[36]. These ideas of different dynamical processes captured in specific regions of k-space should be investigated to find the optimal trade-offs between spatial and temporal resolution for time-resolved imaging.

There are several motion signal surrogates which can be acquired during an MRI examination[37]. Some examples of these are k-space navigator[38] or optical camera[39] based. One promising method is the use of a pilot tone which can supply a high temporal resolution motion sigal[40,41]. Leveraging these motion signal into a low-rank motion field reconstruction such as in the temporal scaling $\Psi$ could further reduce the number of unknowns and improve reconstruction quality.

One drawback of our approach is that it is challenging to separate the motion dynamics from the image reconstruction and only capture them in the motion estimation step. As seen in figure 5, residual motion was captured in the sparse component. There are cases in which complete separation is difficult and fine tuning of the regularization weights and motion field parameters is needed. The use of structure guided total variation (SGTV) in medical imaging registration and reconstruction is promising[42,43]. The idea exploits structural similarity between two images, which is what is desired in the L+S components. Conventional L+S methods try to resolve the motion, hence discouraging the use of SGTV. However, with the proposed method, the L+S decomposition is performed on motion corrected data, which suits the SGTV modelling.

Another drawback of the method is the reconstruction time taken by the current implementation and computing hardware. It is possible to extend the experiments to 3D. However, this increases the reconstruction time and memory required immensely. There are several improvements that can be made. One is simply to use GPUs for speed. Another is to use efficient cartesian sampling for dynamic MRI. A trajectory candidate for this is the stack-of-stars where each blade is the Fourier transform of the imaging objected rotated and a 3D cartesian problem, this would skip the computational demanding interpolation step to a non-cartesian k-space. Finally, with the successes of stochastic gradient descent in deep learning and Extreme MRI[11], similar ideas integrated to CMR-MOTUS would be interesting to explore.

For standard cine imaging, the proposed CMR-MOTUS framework needs a clinical study to confirm the viability of the reconstructions in improving diagnosis. The clinical impact that CMR-MOTUS reconstructions have on perfusion analysis was not investigated, as this work focused on the technical feasibility of the framework. The method uses free-running acquisitions which improves patient comfort and reduces the time necessary to prepare patients since no ECG is required. Additionally, common ECG artifacts such as miss triggering or irregular heartbeats are avoided. Since the cardiac motion fields could be disentangled from the image reconstruction, they might be valuable for determining myocardial strain. Finally, exploring the extension of motion correcting free-running quantitative mapping sequences would improve the clinical utility of the framework.

# Conclusion

In conclusion, CMR-MOTUS offers a significant advancement in capturing cardiac and respiratory motion, as well as contrast dynamics, in first-pass myocardial perfusion MRI without the need for ECG-gating and breath holds. By leveraging a joint-reconstruction approach with the extended MR-MOTUS model and a motion-corrected L+S image model, the framework successfully disentangles different motion and contrast components, providing a more efficient and flexible acquisition method. Despite some challenges, such as the separation of motion dynamics from image reconstruction and the increased reconstruction time for 3D experiments, the proposed method shows promise in improving patient comfort, reducing preparation time, and potentially enhancing clinical utility. Future exploration of multi-resolution motion models, efficient sampling techniques, and integration with deep learning methods could further optimize the framework and its applications in cardiac MRI.